\documentclass[12pt]{iopart}
\usepackage{amsfonts}

\newcommand{\ga}{\gamma}
\newcommand{\la}{\lambda}
\newcommand{\om}{\otimes}

\newcommand{\dl}{\delta}
\newcommand{\ba}{\begin{array}{l}}
\newcommand{\ea}{\end{array}}
\newtheorem{theorem}{Theorem}
\newtheorem{lemma}{Lemma}
\newtheorem{definition}{Definition}
\newtheorem{remark}{Remark}

\begin{document}

\title[]
      {Quantum stochastic equation for a test
      particle interacting with a dilute Bose gas}

\author{A.N. Pechen\footnote[1]{permanent address:
       Steklov Mathematical Institute of Russian Academy of
       Sciences, Gubkin St.~8, 119991, Moscow, Russia}}

\address{Centro Vito Volterra, Universita di Roma Tor Vergata 00133, Roma,
Italy}

\ead{\mailto{pechen@mi.ras.ru}}

\begin{abstract}
We use the stochastic limit method to study long time quantum dynamics of a test particle interacting with a
dilute Bose gas. The case of arbitrary form-factors and an arbitrary, not necessarily equilibrium, quasifree low
density state of the Bose gas is considered. Starting from microscopic dynamics we derive in the low density limit
a quantum white noise equation for the evolution operator. This equation is equivalent to a quantum stochastic
equation driven by a quantum Poisson process with intensity $S-1$, where $S$ is the one-particle $S$ matrix. The
novelty of our approach is that the equations are derived directly in terms of correlators, without use of a
Fock-antiFock (or Gel'fand-Naimark-Segal) representation. Advantages of our approach are the simplicity of
derivation of the limiting equation and that the algebra of the master fields and the Ito table do not depend on
the initial state of the Bose gas. The notion of a causal state is introduced. We construct master fields (white
noise and number operators) describing the dynamics in the low density limit and prove the convergence of
chronological (causal) correlators of the field operators to correlators of the master fields in the causal state.
\end{abstract}

\newpage

\section{INTRODUCTION}

The fundamental equations in quantum theory are the Heisenberg and Schr\" odinger equations. However, it is a very
difficult problem to solve explicitly these equations for realistic physical models and one uses various
approximations or limiting procedures such as weak coupling, low density, and hydrodynamical limits. These scaling
limits describe the long time behavior of physical systems in different physical regimes.

One of the powerful methods to study the long time behavior in quantum theory is the stochastic limit method
developed by Accardi, Lu and Volovich~\cite{ALV}. Many interesting physical models have been investigated by using
this method. In particular, it has been applied to study the long time quantum dynamics of a system interacting
with a reservoir in the case of a weak interaction between the system and reservoir, i.e. in the weak coupling
limit. It was applied to study the spin-boson model~\cite{AKV}, polaron model and nonrelativistic quantum
electrodynamics~\cite{AKV1,AKV2}, quantum Hall effect~\cite{AB}, relations between Hepp-Lieb and Alli-Sewell laser
models~\cite{B}, bifurcation phenomenon in a spin relaxation~\cite{I}, etc.

An important problem is to study the long time dynamics of a quantum system interacting with a reservoir in the
case the interaction is not weak but the density of particles of the reservoir is small, i.e. in the low density
limit. To describe a quantum physical model to which the low density limit can be applied let us consider an
$N$-level atom (test particle) immersed in a free gas whose molecules can collide with the atom; the gas is
supposed to be very dilute. Then the reduced time evolution for the atom will be Markovian, since the
characteristic time $t_S$ for appreciable action of the surroundings on the atom (time between collisions) is much
larger than the characteristic time $t_R$ for relaxation of correlations in the surroundings. The dynamics of the
$N$-level atom interacting with the free gas should converge, in the low density limit, to the solution of a
quantum stochastic differential equation driven by quantum Poisson noise. The quantum Poisson process, introduced
by Hudson and Parthasarathy\,\cite{HP} (for a description of the quantum Poisson process see also
Kumerrer\,\cite{kumerrer}), should arise naturally in the low density limit, as conjectured by Frigerio and
Maassen\,\cite{FrMa} and later by Frigerio and Alicki\,\cite{FA}. For a general survey of quantum stochastic
calculus we refer to the review by Attal\,\cite{A}.

The quantum stochastic equation for the low density limit was derived by Accardi and Lu~\cite{AcLu}--\cite{L}
using perturbation series for the evolution operator. A nonperturbative white noise approach for the investigation
of dynamics in the low density limit is developed in~\cite{APV1,APV2}, where the mathematical procedure, the so
called stochastic golden rule for the low density limit, was formulated. This derivation uses the white noise
technique developed for the case of weak coupling limit by Accardi, Lu and Volovich\,\cite{ALV}. The approach to
derivation of the stochastic equations in~\cite{AcLu}--\cite{APV2} is based on use of the Fock-antiFock
(Gel'fand-Naimark-Segal, or GNS) representation for the canonical commutation relations (CCR) algebra of the Bose
gas. The approach of the present paper does not use the Fock-antiFock representation.

We study the low density limit for an $N$-level atom (test particle) interacting with a Bose gas. Starting from
microscopic quantum dynamics we derive quantum white noise and quantum stochastic differential equations for the
limiting evolution operator. A useful tool is the energy representation introduced in~\cite{APV1,APV2} where the
case of orthogonal formfactors was considered. In the present paper we consider the case of arbitrary formfactors
and an arbitrary, not necessarily equilibrium, quasifree low density state of the reservoir. To each initial low
density state of the Bose gas we associate in the low density limit a special "state" (which is called a causal
state) on the limiting master field algebra. We prove the convergence of time-ordered (or causal) correlators of
the initial Bose field to the correlators of master fields (which are number operators constructed from some white
noise operators) in these causal states. These states are determined by the diagrams which give nontrivial
contribution to the limit. The leading diagrams can be interpreted as a new statistics arising in the low density
limit (new statistics arising in the weak coupling limit is discussed in~\cite{ALV,AAV}).

One of the main results of the paper is that the dynamics in the
low density limit is given by the solution of a quantum white
noise equation, which is equivalent to the quantum stochastic
equation
\begin{equation}\label{1}
\rmd U_t=\rmd N_t(S-1)U_t
\end{equation}
where $U_t$ is the evolution operator at time $t$ describing the limiting dynamics, $S$ is the one-particle $S$
matrix describing scattering of the test particle on one particle of the reservoir, and $N_t(S-1)$ is the quantum
Poisson (number, gauge) process with intensity $S-1$. The equation describes the evolution of the total
system+reservoir and can be applied, in particular, to the important problem of derivation of the linear quantum
Boltzmann equation describing the irreversible reduced dynamics of the test particle in the low density limit.
Such an equation for the reduced density matrix can be easily obtained from the quantum Langevin equation, which
can be derived by using the quantum stochastic differential equation and quantum Ito table (see sect.~\ref{s7})
for stochastic differential ${\rm d}N_t$ (for a derivation of the quantum Langevin equation see~\cite{APV2}).
However, in the present paper we are mainly concentrated on further understanding in what sense the Poisson
process is an approximation of the usual quantum field (Theorem~(\ref{mastfi2})) and in mechanism through which
the quantum stochastic equation arises as a limit of the usual Hamiltonian equation.

In order to describe the objects appearing in~(\ref{1}) let us introduce two Hilbert spaces ${\cal H}_{\rm S}$ and
$\cal H$, which are called in this context the system and one-particle reservoir Hilbert spaces, and the Fock
space $\Gamma(L^2(\mathbb R_+;{\cal H}))$ over the Hilbert space of square-integrable measurable vector-valued
functions from $\mathbb R_+=[0,\infty)$ to $\cal H$. With these notations the solution of the equation is a family
of operators $U_t; t\ge 0$ in ${\cal H}_{\rm S}\otimes\Gamma(L^2(\mathbb R_+;{\cal H}))$ (adapted process); $S$ is
a unitary operator in ${\cal H}_{\rm S}\otimes{\cal H}$, which is explicitly defined in section~\ref{7}.

Let us introduce the notion of a Poisson process. Let $X$ be a self-adjoint operator in a Hilbert space $\cal K$
and $\Psi(f)$ the normalized coherent vector in the Fock space $\Gamma({\cal K})$ with test function $f\in{\cal
K}$. The {\it number operator} is the generator of one-parameter unitary group $\Gamma(e^{i\la X})$ characterized
by
\[
\Gamma(e^{i\la X})\Psi(f)=\Psi(e^{i\la X}f); \qquad\la\in\mathbb R
\]
The number operator is characterized by the property
\[
\langle\Psi(f),N(X)\Psi(g)\rangle=\langle f,Xg\rangle\langle\Psi(f),\Psi(g)\rangle
\]
The definition of $N(X)$ is extended by complex linearity to any bounded operator $X$ on $\cal K$. Let us consider
$\cal K$ of the form $L^2(\mathbb R_+;{\cal H})\cong L^2(\mathbb R_+)\otimes{\cal H}$. For any bounded operators
$X_0\in B({\cal H}_{\rm S})$, $X_1\in B({\cal H})$, and for any $t\ge 0$ we define $N_t(X_0\otimes
X_1):=X_0\otimes N(\chi_{[0,t]}\otimes X_1)$, extend this definition by linearity to any bounded operator $K$ in
${\cal H}_{\rm S}\otimes{\cal H}$, and call the family ${N_t(K); t\ge 0}$ of operators in ${\cal H}_{\rm
S}\otimes\Gamma(L^2(\mathbb R_+;{\cal H}))$ as {\it quantum Poisson process with intensity $K$}. The existence and
uniqueness of the solution of the equation in this case follows from the general theory of quantum stochastic
differential equations. Moreover, unitarity of $S$ leads to the conclusion that for each $t\ge 0$ $U_t$ is a
unitary operator (see Lemma~\ref{unitarity}).

For the vacuum state of the reservoir (zero density) such an equation was derived by Accardi and Lu\,\cite{AL2}.
In the present paper we derive this equation for an arbitrary quasifree initial state of the Bose gas. The main
feature of the present paper is that the stochastic equations are derived directly in terms of correlators,
without use of a Fock-antiFock (or GNS) representation. This simplifies the derivation of the limiting quantum
white noise equation and allows us to express the intensity of the quantum Poisson process directly in terms of
one-particle $S$-matrix. In our approach the limiting equation, the algebra of the master fields, and the Ito
table do not depend on the initial state of the Bose gas.

We obtain that the dynamics of the compound system in the low
density limit is described by:

1) the solution of quantum white noise equation~(\ref{normordeq1}) or equivalently, quantum stochastic
differential equation in forms~(\ref{1}),~(\ref{qsde1}) and

2) the family of causal states $\varphi_L$ on the algebra of master fields.

The reduced dynamics of the system (test particle) in the low density limit for the model under consideration,
with completely different methods, based on a quantum Bogoliubov-Born-Green-Kirkwood-Yvon (BBGKY) hierarchy, has
been investigated by D\"umcke\,\cite{dumcke}, where it is proved that, under some conditions, the reduced dynamics
is given by a quantum Markovian semigroup.

In the approach of the present paper the reduced dynamics can be
easily derived from the solution of the limiting quantum
stochastic differential equation. Namely, the limiting evolution
operator $U_t$ and the limit state $\varphi_L$ determine the
reduced dynamics by
\[
T_t(X)=\varphi_L(U^+_t(X\otimes 1)U_t),
\]
where $X$ is any system observable (bounded operator in ${\cal H}_{\rm S}$), $\varphi_L(\cdot)$ denotes partial
expectation, and $T_t$ is the limiting semigroup. This equality shows that $U_t$ is a stochastic dilation of the
limiting Markovian semigroup. Using the quantum Ito table for stochastic differential ${\rm d}N_t$ one can derive
a quantum Langevin equation for the quantity $U^+_t(X\otimes 1)U_t$. Then taking partial expectation one gets an
equation for $T_t(X)$; in particular, one can get the generator of the semigroup. This is a general feature of the
white noise approach: one at first obtains the equation for the evolution operator of the total system and then
gets the reduced dynamics of the test particle. Let us note that although quantum stochastic equations, which are
derived in~\cite{AcLu,APV2}, are different from~(\ref{1}) they give the same reduced dynamics.

The low density limit can be applied to the model of a test particle moving through an environment of randomly
placed, infinitely heavy scatterers (Lorentz gas) (see the review of Spohn~\cite{spohn}). In the Boltzmann--Grad
limit successive collisions become independent and the averaged over the positions of the scatterers the position
and velocity distribution of the particle converges to the solution of the linear Boltzmann equation. An advantage
of the stochastic limit method is that it allows us to derive equations not only for averaged over reservoir
degrees of freedom dynamics of the test particle but for the total system+reservoir. For a rigorous treatment of a
classical Lorentz gas we refer to~\cite{lanford}--\cite{BBS}. The convergence results and derivation of the linear
Boltzmann equation for a quantum Lorentz gas in the low density and weak coupling limits are presented
in~\cite{EPT,EY1}. The Coulomb gas at low density is considered in~\cite{CLY}.

The hydrodynamical limit is described by the Euler equation. In~\cite{NY} the Euler equation for fermions in the
hydrodynamical limit is derived under some assumptions.

Let us describe the plan of the paper. In section~3 we construct the master fields, which are number operators
acting in some Hilbert space, and the limit causal states on the master field's algebra. We prove that the
time-ordered (or causal) correlators of the free evolution of the initial field converge in the low density limit
to the correlators of the master field in these causal states. In section~4 the stochastic Schr\"odinger equation
which describes the dynamics in the low density limit is derived. In section~5 we bring this equation to the
causally normally ordered form. This form is convenient for study the reduced dynamics of the system. In section~6
the expressions for the one-particle $S$-matrix and $T$-operator are given. In section~7 quantum stochastic
differential equation~(\ref{1}) for the limiting evolution operator is derived.

\section{AN ATOM INTERACTING WITH A DILUTE BOSE GAS}
Let us explain our notations. We consider a quantum system (test particle) interacting with a boson reservoir
(heat bath). Let ${\cal H}_{\rm S} $ be the Hilbert space of the system. For example, for an $N$-level atom ${\cal
H}_{\rm S} = \mathbb C^N$. The system Hamiltonian $H_{\rm S}$ is a self-adjoint operator in ${\cal H}_{\rm S}$.
The reservoir is described by the boson Fock space $\Gamma({\cal H})$ over the one particle Hilbert space ${\cal
H} = L^2(\mathbb R^d)$ (with scalar product $\langle\cdot,\cdot\rangle$), where $d=3$ in the physical case.
Moreover, the Hamiltonian of the reservoir is given by $H_{\rm R}:=\rmd\Gamma(H_1)$ (the second quantization of
the one-particle Hamiltonian $ H_1 $) and the total Hamiltonian $H_{\rm tot}$ of the compound system is given by a
self--adjoint operator on the total Hilbert space ${\cal H}_{\rm S}\otimes \Gamma({\cal H})$:
\[
 H_{\rm tot}:=H_{\rm free}+H_{\rm int}=H_{\rm S}\otimes1+1\otimes H_{\rm R}+H_{\rm int}.
\]
Here $H_{\rm int}$ is the interaction Hamiltonian between the system and reservoir. The one-particle Hamiltonian
$H_1$ is the operator of multiplication by some real-valued function $\omega(k)$. The interaction Hamiltonian will
be assumed to have the following form:
\[
 H_{\rm int}:= \rmi(D\otimes A^+(g_0)A(g_1)-D^+\otimes
 A^+(g_1)A(g_0))
\]
where $ D $ is a bounded operator in $ {\cal H}_S $, $ D\in {\bf B}({\cal H}_S) $; $A(g_n)$ and $ A^+(g_n) $,
$n=0,1$, are annihilation and creation operators, and $ g_0, g_1\in {\cal H}$ are formfactors describing the
interaction of the system with the reservoir. This Hamiltonian describes scattering of particles of the Bose gas
on the test particle and can be obtained by quantization of the classical interaction potential between particles
of two different types with an infinite number of particles of one type (particles of the gas) and finite number
of particles of the second type (test particles). This Hamiltonian preserves the particle number of the reservoir,
and therefore the particles of the reservoir are only scattered on the test particle and not created or destroyed.
Such a Hamiltonian was considered by Davies~\cite{davies} in the analysis of the weak coupling limit.

The initial state of the compound system is supposed to be factorized:
\[
 \rho = \rho_{\rm S}\otimes\varphi_{L,\xi}.
\]
Here $\rho_{\rm S}$ is an arbitrary density matrix of the system and the initial state of the reservoir
$\varphi_{L,\xi}$ is the gauge invariant mean zero Gaussian state, characterized by
\begin{equation}\label{state}
 \varphi_{L,\xi}(A^+(f)A(g))=\xi\left\langle g,\frac{L}{1-\xi L}f\right\rangle
\end{equation}
for each $f,g\in{\cal H}$. Here $\xi>0$ is a small positive number and $L$ is a bounded positive operator in
${\cal H}$ commuting with $S_t$ (an operator of multiplication by some function $L(k)$). In the case
$L=\rme^{-\beta H_1}$, so that $L(k)=\rme^{-\beta\omega(k)}$, where $\beta>0$ is a positive number, the state
$\varphi_{L,\xi}$ is just the Gibbs state, at inverse temperature $\beta$ and fugacity $\xi$, of the free
evolution. The fugacity $\xi=\rme^{\beta\mu}$; $\mu$ is the chemical potential.

The dynamics of the total system is determined by the evolution operator which in interaction representation has
the form:
\[
 U(t):=\rme^{\rmi tH_{\rm free}}\rme^{-\rmi tH_{\rm tot}}.
\]
It satisfies the differential equation
\[
 \frac{dU(t)}{dt}=-\rmi H_{\rm int}(t)U(t),
\]
where the quantity $H_{\rm int}(t)$ will be
called the evolved interaction and defined as
\[
 H_{\rm int}(t)=\rme^{\rmi tH_{\rm free}} H_{\rm int}\rme^{-\rmi tH_{\rm free}}.
\]
The iterated series for the evolution operator is
\begin{equation}\label{eqU111}
U(t)=1+\sum\limits_{n=1}^\infty(-\rmi)^n\int\limits_0^tdt_1\dots\int\limits_0^{t_{n-1}}dt_n
H_{\rm int}(t_1)\dots H_{\rm int}(t_n)
\end{equation}
With the notations
\[
 S_t:=\rme^{\rmi tH_1}\ ,\qquad D(t):=\rme^{\rmi tH_{\rm S}}D\rme^{-\rmi tH_{\rm S}}
\]
the evolved interaction can be written in the form
\begin{equation}
 H_{\rm int}(t):=
 \rmi(D(t)\otimes A^+(S_tg_0)A(S_tg_1)-D^+(t)\otimes A^+(S_tg_1)A(S_tg_0)).\label{Hint}
\end{equation}
We assume the rotating wave approximation
\[
 \rme^{\rmi tH_{\rm S}}D\rme^{-\rmi tH_{\rm S}} = D,
\]
although generalization to the case of arbitrary $D$ is not
difficult.

We study the dynamics generated by the Hamiltonian~(\ref{Hint}) in the low density limit: $n\to 0$, $t\sim 1/n$
($n$ is the density of particles of the reservoir). The density of particles with momentum $k$ in the state
$\varphi_{L,\xi}$ is equal to
\[
\frac{\xi L(k)}{1-\xi L(k)}
\]
and goes to zero as $\xi\to 0$. Therefore the limit $n\to 0$,
$t\sim 1/n$ is equivalent to the limit $\xi\to 0$, $t\sim 1/\xi$.

Let us consider the time rescaling $ t\to t/\xi $ so that $U(t)\to
U(t/\xi)$. With the notation
\begin{equation}\label{2}
 N_{f,g,\xi}(t)=\frac{1}{\xi}A^+(S_{t/\xi}f)
 A(S_{t/\xi}g)
\end{equation}
for any $f,g\in{\cal H}$, the equation for the evolution operator
$U(t/\xi)$ becomes
\begin{equation}\label{eqev}
 \frac{\rmd U(t/\xi)}{\rmd t}=(D\otimes N_{g_0,g_1,\xi}(t)-D^+\otimes
 N_{g_1,g_0,\xi}(t))U(t/\xi)
\end{equation}

The reduced dynamics of any test particle's observable $X$ in the
low density limit is defined by the limit
\[
\lim\limits_{\xi\to 0}\varphi_{L,\xi}(U^+(t/\xi)(X\otimes
1)U(t/\xi))
\]
where $\varphi_{L,\xi}(\cdot)$ denotes partial expectation. In~\cite{dumcke} it was proved that, under some
conditions, the limit exists in a small time interval and is equal to $T_t(X)$, where $\{T_t; t\ge 0\}$ is a
quantum Markovian semigroup.  The dynamics of the reduced density matrix $\rho_{\rm S}(t)$ is determined through
the duality ${\rm Tr}(\rho_{\rm S}T_t(X))={\rm Tr}(\rho_{\rm S}(t)X)$. As was mentioned in the Introduction, in
the approach of the present paper the limiting semigroup can be obtained by using the solution $U_t$ of the
quantum stochastic equation as
\[
T_t(X)=\varphi_L(U^+_t(X\otimes 1)U_t)
\]
and the generator of the semigroup can be easily derived from quantum Langevin equation. The limiting semigroup
can be obtained also from quantum Langevin equation in~\cite{APV2}, which is based on a quantum stochastic
equation similar to~(\ref{1}) but much more complicated.

The first step to study the low density limit of the model is to
find the limit of the field $N_{f,g,\xi}(t)$. This limit we call
master fields or number operators.

\section{THE MASTER FIELDS AND THE LIMIT STATES}

In this section we construct the algebra of the master fields arising in the low density limit and the limit
causal states on this algebra. We prove (Theorem~1) that time-ordered correlators of initial fields~(\ref{2})
converge in the low density limit to correlators of number operators constructed from some white noise operators.
Theorem~2 states a useful factorization property of the limiting causal states.

It is convenient to use the "projections"
\[
 P_E:=\frac{1}{2\pi}\int\limits_{-\infty}^{\infty}\rmd tS_t\rme^{-\rmi tE}=
 \dl(H_1-E)
\]
with the properties
\[
 P_EP_{E'}=\dl(E-E')P_E,\qquad P^*_E=P_E,\qquad S_t=\int\rmd EP_E\rme^{\rmi tE}
\]
For the $\dl$-function of a self-adjoint operator cf. Definition
(1.2.1) in~\cite{ALV}.

Let us construct the master space (which is Fock space over some
Hilbert space) and master fields. For a given Hilbert space ${\cal
H}$ and a self-adjoint operator $H_1$ in ${\cal H}$ we define the
Hilbert space ${\cal X}_{{\cal H},H_1}$ as the completion of the
quotient of the set
\[
\left\{F: \mathbb R\to{\cal H}\,\,\,{\rm s.t.}\,\,\, ||F||^2:=2\pi\int\rmd E\langle
F(E),P_EF(E)\rangle<\infty\right\}
\]
with respect to the zero-norm elements. The inner product in
${\cal X}_{{\cal H},H_1}$ is defined as
\[
\langle F,G\rangle=2\pi\int\rmd E\langle F(E),P_EG(E)\rangle.
\]
We denote by $B^+_f(E,t),\, B_g(E',t')$ time-energy white noise creation and annihilation operators acting in the
symmetric Fock space $\Gamma(L^2(\mathbb R_+,{\cal X}_{{\cal H},H_1}))$ where $L^2(\mathbb R_+,{\cal X}_{{\cal
H},H_1})$ is the Hilbert space of square integrable functions $f: \mathbb R_+\to{\cal X}_{{\cal H},H_1}$. These
operators (operator-valued distributions) satisfy the canonical commutation relations
\begin{equation}\label{ccrB0}
[B_g(E,t),\, B^+_f(E',t')]=\dl(t'-t)\dl(E'-E)\tilde\gamma_{g,f}(E)
\end{equation}
and causal commutation relations
\begin{equation}\label{ccrB}
[B_g(E,t),\, B^+_f(E',t')]=\dl_+(t'-t)\dl(E'-E)\gamma_{g,f}(E)
\end{equation}
where $\dl_+(t'-t)$ is the causal $\dl$-function and
\[
\gamma_{g,f}(E)=\int dE'\frac{\langle g,P_{E'}f\rangle}{\rmi(E'-E-i0)}
\]
\[
\tilde\gamma_{g,f}(E)= 2\pi\langle g,P_Ef\rangle
\]
In the Appendix we review the definition of the causal $\dl$-function; for a detailed discussion of distributions
over the simplex and the meaning of two different commutators~(\ref{ccrB0}) and~(\ref{ccrB}) for the same
operators we refer to Sect.~7 in~\cite{ALV}. These operators are called time-energy quantum white noise due to the
presence of $\dl(t'-t)\dl(E'-E)$ in~(\ref{ccrB0}).

For any positive bounded operator $L$ in $\cal H$ we define the causal gauge-invariant mean-zero Gaussian state
$\varphi_L$ by the properties (\ref{prop1})-(\ref{state1}):
\begin{equation}\label{prop1}{\rm for\,\, n=2k}\qquad
\varphi_L(B^{\epsilon_1}_1\dots B^{\epsilon_n}_n)=
\sum\varphi_L(B^{\epsilon_{i_1}}_{i_1}B^{\epsilon_{j_1}}_{j_1})
\dots\varphi_L(B^{\epsilon_{i_k}}_{i_k}B^{\epsilon_{j_k}}_{j_k})
\end{equation}
where the sum is taken over all permutations of the set
$(1,\dots,2k)$ such that $i_\alpha< j_\alpha$, $\alpha=1,\dots,k$,
$i_1<i_2<\dots<i_k$;
$B^{\epsilon_m}_m:=B^{\epsilon_m}_{f_m}(E_m,t_m)$ for
$m=1,\dots,n$, are time-energy quantum white noise operators with
causal commutation relations~(\ref{ccrB}), and $\epsilon_m$ means
either creation or annihilation operator;
\begin{equation}{\rm for\,\, n=2k+1}\qquad
\varphi_L(B^{\epsilon_1}_1\dots B^{\epsilon_n}_n)=0
\end{equation}
\begin{equation}
\varphi_L(B_f(E,t)B_g(E',t'))= \varphi_L(B^+_f(E,t)B^+_g(E',t'))=0
\end{equation}
\begin{equation}\label{state1}
\varphi_L(B^+_f(E,t)B_g(E',t'))=\chi_{[0,t]}(t')\langle g,P_ELf\rangle
\end{equation}

Notice that the "state" $\varphi_L$ does not satisfy the positivity condition. This is a well-known situation for
the weak coupling limit (see~\cite{ALV}) and is due to the fact that we work with time-ordered, or causal
correlators. Therefore it is natural to call such "states" causal states.

\begin{definition}
Causal time-energy white noise is a pair $(B_f(E,t),\varphi_L)$, where $B_f(E,t)$ satisfy the causal commutation
relations~(\ref{ccrB}) and $\varphi_L$ is a causal gauge-invariant mean-zero Gaussian state characterized
by~(\ref{prop1})-(\ref{state1}).
\end{definition}
Using the operators $B^+_f(E,t)$, $B_g(E,t)$ we define the number
operators as
\begin{equation}\label{defNfg}
N_{f,g}(t)=\int dEB^+_f(E,t)B_g(E,t)
\end{equation}

Finally, for a given Hilbert space ${\cal H}$ and a self-adjoint operator $H_1$ we have the following objects: for
any $\xi>0$ the family of operators $N_{f,g,\xi}(t)$ defined by~(\ref{2}) together with the gauge-invariant
quasifree mean-zero Gaussian state $\varphi_{L,\xi}$ and the number operators $N_{f,g}(t)$ together with the
causal state $\varphi_L$.

The following theorem describes the relation between these objects
and states the master field in the low density limit.
\begin{theorem}\label{mastfi2}
There exists causal time-energy white noise $(B_f(E,t),\varphi_L)$ such that $\forall n\in\mathbb N$
\[
\lim\limits_{\xi\to 0}\varphi_{L,\xi}(N_{f_1,g_1,\xi}(t_1)\dots
N_{f_n,g_n,\xi}(t_n))=\varphi_L(N_{f_1,g_1}(t_1)\dots
N_{f_n,g_n}(t_n))
\]
where the equality is understood in the sense of distributions
over simplex $t_1\ge t_2\ge\dots \ge t_n\ge0$. The limit causal
state $\varphi_L$ is characterized by~(\ref{prop1})-(\ref{state1})
and the number operators are defined by~(\ref{defNfg}).
\end{theorem}
\begin{remark}
This convergence is called convergence in the sense of time-ordered correlators. The fact that we use the
distributions over simplex is motivated by iterated series~(\ref{eqU111}) for the evolution operator.
\end{remark}
{\bf Proof.} Notice that
\[
 N_{f,g,\xi}(t)=\int \rmd EN_{f,g,\xi}(E,t)
\]
where
\[
 N_{f,g,\xi}(E,t):=\frac{\rme^{\rmi
 tE/\xi}}{\xi}A^+(P_Ef)A(S_{t/\xi}g)
\]
Therefore
\[\fl
\varphi_{L,\xi}(N_{f_1,g_1,\xi}(t_1)\dots N_{f_n,g_n,\xi}(t_n))=
\int\rmd E_1\dots\rmd
E_n\varphi_{L,\xi}(N_{f_1,g_1,\xi}(E_1,t_1)\dots
N_{f_n,g_n,\xi}(E_n,t_n))
\]
Let us denote for shortness of notation for $l=1,\dots, n$,
\[
A^+_l:=\frac{\rme^{\rmi t_lE_l/\xi}}{\sqrt{\xi}}A^+(P_{E_l}f_l);
\qquad A_l:=\frac{1}{\sqrt{\xi}}A(S_{t_l/\xi}g_l)
\]
In this notation
\begin{equation}\label{eq7}
\varphi_{L,\xi}(N_{f_1,g_1,\xi}(E_1,t_1)\dots
N_{f_n,g_n,\xi}(E_n,t_n)) = \varphi_{L,\xi}(A^+_1A_1\dots
A^+_nA_n)
\end{equation}

The state $\varphi_{L,\xi}$ is a gauge-invariant mean-zero Gaussian state. Therefore~(\ref{eq7}) is equal to the
sum of terms of the form
\begin{equation}\label{eq8}
\varphi_{L,\xi}(A^+_{i_1}A_{j_1})\dots\varphi_{L,\xi}(A^+_{i_k}A_{j_k})
\varphi_{L,\xi}(A_{j_{k+1}}A^+_{i_{k+1}})\dots\varphi_{L,\xi}(A_{j_n}A^+_{i_n})
\end{equation}
where $k=1,\dots, n$, $1=i_1<i_2<\dots<i_k$, $j_{k+1}<\dots<j_n$, $i_l\le j_l$ for $l=1,\dots,k$ and $j_l<i_l$ for
$l=k+1,\dots,n$. We say that~(\ref{eq8}) corresponds to a nonconnected diagram if there exists $m\in\{1,\dots,n\}$
such that $i_l\le m\Leftrightarrow j_l\le m$. Otherwise we say that~(\ref{eq8}) corresponds to a connected
diagram.

Let us prove that all the connected diagrams except only one
corresponding to the case $k=1$ are equal to zero in the limit.
One can write~(\ref{eq8}) as
\begin{eqnarray}\fl
\frac{1}{\xi^n}\exp\Bigl\{i[(t_1-t_{j_1})E_1+
\dots+(t_{i_n}-t_{j_n})E_{i_n}]/\xi\Bigr\}
\Bigl(\xi^kF(E)+O(\xi^{k+1})\Bigr)\nonumber\\
\fl=\frac{1}{\xi^n}\exp\Bigl\{i[t_n(E_n-E_{\alpha_n})+
\dots+t_1(E_1-E_{\alpha_1})]/\xi\Bigr\}\Bigl(\xi^kF(E)+O(\xi^{k+1})\Bigr)\nonumber\\
\fl=\frac{1}{\xi^{n-1}}\exp\Bigl\{i[(t_n-t_{n-1})\omega_n(E)+
\dots+(t_2-t_1)\omega_2(E)]/\xi\Bigr\}\Bigl(\xi^{k-1}F(E)+O(\xi^k)\Bigr)\nonumber\\
=\frac{\rme^{\rmi(t_n-t_{n-1})\omega_n(E)/\xi}}{\xi}\dots
\frac{\rme^{\rmi(t_2-t_1)\omega_2(E)/\xi}}{\xi}
\Bigl(\xi^{k-1}F(E)+O(\xi^k)\Bigr)\label{eq9}
\end{eqnarray}
where $(\alpha_1,\dots,\alpha_n)$ is the permutation of the set
$(1,\dots,n)$,
$\omega_l(E)=E_n+\dots+E_l-E_{\alpha_n}-\dots-E_{\alpha_l}$ for
$l=2,\dots,n$ and
\[
F(E)=\prod\limits_{l=1}^k\langle g_{j_l},P_{E_l}Lf_{i_l}\rangle\prod\limits_{l=k+1}^n\langle
g_{j_l},P_{E_{i_l}}f_{i_l}\rangle
\]
Notice that for a connected diagram all the functions
$\omega_l(E)$ are not identically zero. In fact, suppose that
$\omega_m(E)\equiv 0$ for some $m\in\{2,\dots,n\}$. In this case
one has the identity
\[
E_m+\dots+E_n\equiv E_{\alpha_m}+\dots+E_{\alpha_n}
\]
(where $E_\alpha, E_{\alpha'}$ for $\alpha\neq\alpha'$ are independent variables) which means that
$(\alpha_m,\dots,\alpha_n)$ is a permutation of the set $\{m,\dots,n\}$ and hence $(\alpha_1,\dots,\alpha_{m-1})$
is a permutation of the set $\{1,\dots,m-1\}$. Let us choose any $l\in\{1,\dots,n\}$ and consider the term
$t_{j_l}(E_{j_l}-E_{i_l})$ in the exponent in the second line of~(\ref{eq9}). If $j_l<m$, then since
$i_l\equiv\alpha_{j_l}$ and $\alpha_{j_l}$ belongs to the set $\{1,\dots,m-1\}$ one has
$i_l\equiv\alpha_{j_l}\in\{1,\dots,m-1\}$, and vice versa if $\alpha_{j_l}\equiv i_l\in\{1,\dots,m-1\}$, then
$j_l\le m-1$. This means that if $\omega_l$ are not identically zero, then~(\ref{eq8}) corresponds to a connected
diagram.

Let us consider the case $k>1$. Then, if~(\ref{eq8}) corresponds to a connected diagram, the functions
$\omega_l(E)$ are not identically zero. In this case, since there exists the limit
\[
\lim\limits_{\xi\to 0}
\frac{\rme^{\rmi(t_l-t_{l-1})\omega_l(E)/\xi}}{\xi} =
\dl_+(t_l-t_{l-1})\frac{1}{\rmi(\omega_l(E)-\rmi 0)}
\]
and the limit of the product of such terms in~(\ref{eq9}), and $k-1>0$, the limit of~(\ref{eq9}) is equal to zero.

Now let us consider the case $k=1$. In this case~(\ref{eq8}) has
the form
\begin{eqnarray}\fl
\varphi_{L,\xi}(A^+_1A_n)\varphi_{L,\xi}(A_1A^+_2)\dots
\varphi_{L,\xi}(A_{n-1}A^+_n)=\nonumber\\
\fl\frac{1}{\xi^n}\exp\Bigl\{i[(t_1-t_n)E_1+(t_2-t_1)E_2+
\dots+(t_n-t_{n-1})E_n]/\xi\Bigr\}
\Bigl(\xi F(E)+O(\xi^2)\Bigr)\nonumber\\
=\frac{\rme^{\rmi(t_n-t_{n-1})\omega_n(E)/\xi}}{\xi}\dots
\frac{\rme^{\rmi(t_2-t_1)\omega_2(E)/\xi}}{\xi}
\Bigl(F(E)+O(\xi)\Bigr)\label{eq12}
\end{eqnarray}
where $\omega_l(E)=E_l-E_1$. Using the limit~(\ref{dlplimit}) one finds that the limit of the right-hand side
(RHS) of~(\ref{eq12}) is equal to
\[\fl
\dl_+(t_2-t_1)\dots\dl_+(t_n-t_{n-1})\langle g_n,P_{E_1}Lf_1\rangle\frac{\langle
g_1,P_{E_2}f_2\rangle}{\rmi(E_2-E_1-\rmi 0)}\dots \frac{\langle g_{n-1},P_{E_n}f_n\rangle}{\rmi(E_n-E_1-\rmi 0)}
\]
After integration over $E_1\dots E_n$ it becomes equal to
\begin{equation}\label{eq13}\fl
\dl_+(t_2-t_1)\dots\dl_+(t_n-t_{n-1})\int\rmd E\langle g_n,P_ELf_1\rangle
\gamma_{g_1,f_2}(E)\dots\gamma_{g_{n-1},f_n}(E)
\end{equation}
This proves that only one connected diagram survives in the limit.

Now let us consider the quantity
\[
\varphi_L(N_{f_1,g_1}(t_1)\dots N_{f_n,g_n}(t_n))
\]
With the notation
\[
B^+_l:=B^+_{f_l}(E_l,t_l);\qquad B_l:=B_{g_l}(E_l,t_l),
\]
it can be written as
\begin{equation}\label{eq14}
\int\rmd E_1\dots\rmd E_n\varphi_L(B^+_1B_1\dots B^+_nB_n)
\end{equation}
Notice that on the simplex $t_1\ge t_2\ge\dots\ge t_n\ge 0$ causal
$\dl$-functions $\dl_+(t_{l+m}-t_l)$ for $m\ge2$ are equal to
zero. Therefore for $m\ge2$ one has
$\varphi_L(B_{t_l}B^+_{t_{l+m}})\propto\dl_+(t_{l+m}-t_l)=0$ and
hence the integrand in~(\ref{eq14}) can be written as
\begin{eqnarray}\fl
\varphi_L(B^+_1B_1\dots B^+_nB_n)=\sum\limits_{k=1}^{n-1}
\varphi_L(B^+_1B_k)\varphi_L(B_1B^+_2)\dots\varphi_L(B_{k-1}B^+_k)
\varphi_L(B^+_{k+1}B_{k+1}\dots B^+_nB_n)\nonumber\\
+\varphi_L(B^+_1B_n)\varphi_L(B_1B^+_2)\dots\varphi_L(B_{n-1}B^+_n)
\end{eqnarray}
The terms in the sum correspond to nonconnected diagrams. The last
term corresponds to a unique nonzero connected diagram. Moreover
\begin{eqnarray}\fl
\int\rmd E_1\dots\rmd
E_n\varphi_L(B^+_1B_n)\varphi_L(B_1B^+_2)\dots\varphi_L(B_{n-1}B^+_n)=
\dl_+(t_2-t_1)\dots\dl_+(t_n-t_{n-1})\nonumber\\
\times\int\rmd E <g_n,P_ELf_1> \gamma_{g_1,f_2}(E)\dots\gamma_{g_{n-1},f_n}(E),\nonumber
\end{eqnarray}
which is equal to~(\ref{eq13}).

For $n=1$ the statement of the theorem is clear. In fact,
\[\fl
\lim\limits_{\xi\to 0}\varphi_{L,\xi}(N_{f,g,\xi}(t)) = \lim\limits_{\xi\to 0}\left\langle g,\frac{L}{1-\xi
L}f\right\rangle =\langle g,Lf\rangle=\int\rmd E\varphi_L(B^+_f(E,t)B_g(E,t))
\]
Then proof of the theorem follows by induction using the fact that
only one connected diagram survives in the limit.
\begin{remark} The fact that in each order of iterated series only one connected diagram survives
in the limit can be interpreted as emergence of a new statistics
(different from Bose) in the low density limit. For a discussion
of new statistic arising in the weak coupling limit we refer
to\,\cite{ALV} (see also\,\cite{AAV}).
\end{remark}

The following theorem is important for investigation of the
limiting white noise equation for the evolution operator.
\begin{theorem}\label{limitstate} The limit state $\varphi_L$ has the following
factorization property: $\forall n\in\mathbb N$,
\begin{eqnarray}
\varphi_L(B^+_f(E,t)N_{f_1,g_1}(t_1)\dots
N_{f_n,g_n}(t_n)B_g(E,t))\nonumber\\
= \varphi_L(B^+_f(E,t)B_g(E,t))\varphi_L(N_{f_1,g_1}(t_1)\dots
N_{f_n,g_n}(t_n))
\end{eqnarray}
where the equality is understood in the sense of distributions
over simplex $t\ge t_1\ge t_2\ge\dots \ge t_n\ge0$.
\end{theorem}
{\bf Proof.} From Gaussianity of the causal state $\varphi_L$
(property~(\ref{prop1})) it follows that
\begin{eqnarray}
 \varphi_L(B^+_f(E,t)N_{f_1,g_1}(t_1)\dots
N_{f_n,g_n}(t_n)B_g(E,t))\nonumber\\
 = \varphi_L(B^+_f(E,t)B_g(E,t))\varphi_L(N_{f_1,g_1}(t_1)\dots
N_{f_n,g_n}(t_n))\nonumber\\
 +\int dE_1\dots
 dE_n\sum\varphi_L(B^+_f(E,t)B_{g_i}(E_i,t_i))\dots
 \varphi_L(B^+_{f_j}(E_j,t_j)B_g(E,t))\nonumber
\end{eqnarray}
The sum is equal to zero since the last multiplier
\[
\varphi_L(B^+_{f_j}(E_j,t_j)B_g(E,t))=\chi_{[0,t_j]}(t)<g,P_{E_j}Lf_j>
\]
is equal to zero almost everywhere on the simplex $t\ge t_1\ge
t_2\ge\dots \ge t_n\ge0$ and hence is equal to zero in the sense
of distributions on the simplex. This proves the theorem.

Theorem~1 allows us to calculate, in particular, the partial expectation of the evolution operator and Heisenberg
evolution of any system observable in the low density limit. In fact, partial expectation of the $n$-th term of
the iterated series for the evolution operator~(\ref{eqU111}) (or equivalent series for Heisenberg evolution of a
system observable) after time rescaling $t\to t/\xi$ includes the quantity
\[
\int\limits_0^t\rmd t_1\dots\int\limits_0^{t_{n-1}}\rmd t_n
\varphi_{L,\xi}(N_{f_1,g_1,\xi}(t_1)\dots N_{f_n,g_n,\xi}(t_n))
\]
(where $f_\alpha,\, g_\alpha$ are equal to $g_0$ or $g_1$). The limit as $\xi\to 0$ of this quantity can be
calculated using Theorem~1. For example, the contribution of the connected diagram is equal to
\begin{eqnarray}
\int\limits_0^t\rmd t_1\int\limits_0^{t_1}\rmd
t_2\dl_+(t_2-t_1)\int\limits_0^{t_2}\rmd t_3\dl_+(t_3-t_2)\dots
\int\limits_0^{t_{n-1}}\rmd t_n\dl_+(t_n-t_{n-1})\nonumber\\
\times\int\rmd E
\langle g_n,P_ELf_1\rangle\gamma_{g_1,f_2}(E)\dots\gamma_{g_{n-1},f_n}(E)\nonumber\\
=t\int\rmd E \langle g_n,P_ELf_1\rangle\gamma_{g_1,f_2}(E)\dots\gamma_{g_{n-1},f_n}(E)\nonumber
\end{eqnarray}
Similarly one can calculate the contribution of nonconnected diagrams (they give terms proportional to higher
orders of $t$). Summing over all orders of the iterated series one can find the reduced dynamics of the system.
But in the present paper we will get the limiting dynamics in a nonperturbative way, without direct summation of
the iterated series. This procedure includes derivation of the white noise equation for the limiting evolution
operator and then bringing this equation to the causally normally ordered form. After that one can easily find,
for example, the reduced dynamics of the system. For the weak coupling limit such a procedure was developed
in\,\cite{ALV}. A nontrivial generalization to the low density limit was developed in\,\cite{APV1,APV2}, where the
derivation is based on the Fock-antiFock representation for the CCR algebra of the Bose field determined by the
state $\varphi_{L,\xi}$. The approach of the present paper does not require a GNS representation and is different
from approach of~\cite{APV1,APV2}.

\section{THE WHITE NOISE SCHR\"ODINGER EQUATION}
In this section we derive, using the results of previous section,
the white noise Schr\"odinger equation for the limiting evolution
operator.

The evolution operator $U(t/\xi)$ satisfies equation~(\ref{eqev}) which can be written as
\[
 \frac{\rmd U(t/\xi)}{\rmd t}=-\rmi H_\xi(t)U(t/\xi),
\]
where
\[
H_\xi(t)=\rmi(D\om N_{g_0,g_1,\xi}(t)-D^+\om N_{g_1,g_0,\xi}(t))
\]
The results of the preceding section allow us to write the limit
as $\xi\to 0$ of the Hamiltonian $H_\xi(t)$. In the
notation~(\ref{defNfg}) the limiting Hamiltonian is the following
operator in $ {\cal H}_S\om\Gamma(L^2(\mathbb R_+,{\cal X}_{{\cal
H},H_1}))$:
\begin{eqnarray}
 H(t)=\rmi(D\om N_{g_0,g_1}(t)-D^+\om N_{g_1,g_0}(t))\nonumber\\
 = \rmi\int\rmd E\Bigl(D\om B^+_{g_0}(E,t)B_{g_1}(E,t)-
 D^+\om B^+_{g_1}(E,t)B_{g_0}(E,t)\Bigr)\label{H}
\end{eqnarray}

The dynamics of the total system (system+reservoir) in the low
density limit $\xi\to 0$ is given by a new evolution operator
$U_t$ which is the solution of the white noise Schr\"odinger
equation
\begin{equation}\label{equ1}
\frac{\rmd U_t}{\rmd t}=-\rmi H(t)U_t,\qquad U_0=1,
\end{equation}
or equivalent integral equation
\begin{equation}\label{equ3}
U_t=1+\int\limits_0^t\rmd t_1\Bigl(D\om N_{g_0,g_1}(t_1)-D^+\om
N_{g_1,g_0}(t_1)\Bigr)U_{t_1}.
\end{equation}

\section{NORMALLY ORDERED FORM OF THE WHITE NOISE EQUATION}
Our next step is to bring the white noise Schr\"odinger equation to the causally normally ordered form
(Theorem~3), i.e., the form in which all annihilation operators are on the right side of the evolution operator
and all creation operators are on the left side. Such a form is convenient for study of the limiting dynamics (see
remark~3 and text after remark). In particular, it can be used for derivation of (linear) Boltzmann equation.

We assume that for each $ E\in {\mathbb R}$, the inverse operators
\[\fl
 T_0(E):=\Bigl(1+\ga_{g_0,g_1}(E)D^+-\ga_{g_1,g_0}(E)D+(\ga_{g_0,g_0}\ga_{g_1,g_1}-
 \ga_{g_1,g_0}\ga_{g_0,g_1})(E)DD^+\Bigr)^{-1}
\]
\[\fl
 T_1(E):=\Bigl(1+\ga_{g_0,g_1}(E)D^+-\ga_{g_1,g_0}(E)D+
 (\ga_{g_0,g_0}\ga_{g_1,g_1}-\ga_{g_1,g_0}\ga_{g_0,g_1})(E)D^+D\Bigr)^{-1}
\]
exist.
\begin{lemma}\label{normordu}
If the evolution operator $U_t$ satisfies~(\ref{equ1}) with $H(t)$ given by~(\ref{H}) then one has
\begin{eqnarray}
 \fl B_{g_0}(E,t)U_t=\ga_{g_0,g_0}(E)T_0(E)DU_tB_{g_1}(E,t)+
 T_0(E)(1-\ga_{g_1,g_0}(E)D)U_tB_{g_0}(E,t)\label{commbf}\\
 \fl B_{g_1}(E,t)U_t=-\ga_{g_1,g_1}(E)T_1(E)D^+U_tB_{g_0}(E,t)+
 T_1(E)(1+\ga_{g_0,g_1}(E)D^+)U_tB_{g_0}(E,t)\label{commbg}
\end{eqnarray}
\end{lemma}
\begin{remark}
Notice that in the RHS of these equalities the annihilation
operators $B_f(E,t)$ are on the right of the evolution operator.
\end{remark}
{\bf Proof.} It follows from~(\ref{ccrB}) and~(\ref{defNfg}) that
\begin{equation}\label{cmrelBN}
[B_{f'}(E,t), N_{f,g}(t_1)]=\dl_+(t_1-t)\ga_{f',f}(E)B_g(E,t)
\end{equation}
Therefore using the integral equation~(\ref{equ3}) for the
evolution operator one gets
\begin{eqnarray}
 \fl B_f(E,t)U_t=[B_f(E,t),U_t]+U_tB_f(E,t)\nonumber\\
 \fl = \int\limits_0^t\rmd t_1
 \Bigl(D\om[B_f(E,t),N_{g_0,g_1}(t_1)]-D^+\om[B_f(E,t),N_{g_1,g_0}(t_1)]
 \Bigr)U_{t_1}+U_tB_f(E,t)\nonumber\\
 =\Bigl(D\ga_{f,g_0}(E)B_{g_1}(E,t)-D^+\ga_{f,g_1}(E)B_{g_0}(E,t)\Bigr)U_t
 +U_tB_f(E,t)\label{commbu1}
\end{eqnarray}
The second equality in~(\ref{commbu1}) holds because, due to the
time consecutive principle
\[
 [B_f(E,t),U_{t_1}]=0 \quad{\rm for}\,\, t_1<t.
\]
In fact, let us consider the quantity
\begin{equation}\label{eqq0}\fl
\int\limits_0^t\rmd t_1[B_f(E,t),U_{t_1}^{(n-1)}]=
(-i)^{n-1}\int\limits_0^t\rmd t_1\dots\int\limits_0^{t_{n-1}}\rmd
t_n [B_f(E,t),H(t_2)\dots H(t_n)]
\end{equation}
where the $n$-th term of the iterated series~(\ref{eqU111}) for
$U_t$ has the form
\[
U_t^{(n)}:=(-i)^n\int\limits_0^t\rmd
t_1\dots\int\limits_0^{t_{n-1}} \rmd t_nH(t_1)\dots H(t_n)
\]
The commutator $[B_f(E,t),H(t_k)]$ proportional to $\dl_+(t_k-t)$,
hence the commutator $ [B_f(E,t),H(t_2)\dots H(t_n)]$ is equal to
zero on the simplex $t\ge t_1\ge t_2\dots\ge t_n\ge 0$ and
therefore~(\ref{eqq0}) is equal to zero.

The third equality in~(\ref{commbu1}) holds since from~(\ref{cmrelBN}) and the definition of causal $\dl$-function
one has
\[
\int\limits_0^t\rmd t_1\dl_+(t_1-t)B_f(E,t_1)U_{t_1}=B_f(E,t)U_t
\]
For a detailed discussion of the time consecutive principle and causal $\dl$-function we refer to\,\cite{ALV}.

After the substitution $f=g_0$ and $f=g_1$ in~(\ref{commbu1}) one gets
\begin{eqnarray}
 \fl B_{g_0}(E,t)U_t=\Bigl(D\ga_{g_0,g_0}(E)B_{g_1}(E,t)
 -D^+\ga_{g_0,g_1}(E)B_{g_0}(E,t)\Bigr)U_t+U_tB_{g_0}(E,t)\nonumber\\
 \fl B_{g_1}(E,t)U_t=\Bigl(D\ga_{g_1,g_0}(E)B_{g_1}(E,t)-
 D^+\ga_{g_1,g_1}(E)B_{g_0}(E,t)\Bigr)U_t+U_tB_{g_1}(E,t)\nonumber
\end{eqnarray}
or equivalently
\begin{eqnarray}
 \fl (1+\ga_{g_0,g_1}(E)D^+)B_{g_0}(E,t)U_t=
 \ga_{g_0,g_0}(E)DB_{g_1}(E,t)U_t+U_tB_{g_0}(E,t)\label{eq1}\\
 \fl (1+\ga_{g_1,g_0}(E)D)B_{g_1}(E,t)U_t=
 -\ga_{g_1,g_1}(E)D^+B_{g_0}(E,t)U_t+U_tB_{g_1}(E,t)\label{eq2}
\end{eqnarray}
After left multiplication of both sides of equality~(\ref{eq1}) by
$(1+\ga_{g_1,g_0}(E)D)$ and both sides of~(\ref{eq2}) by
$\ga_{g_0,g_0}(E)D$ one gets
\begin{eqnarray}
 \fl (1+\ga_{g_1,g_0}(E)D)(1+\ga_{g_0,g_1}(E)D^+)B_{g_0}(E,t)U_t=
 \ga_{g_0,g_0}(E)D(1+\ga_{g_1,g_0}(E)D)B_{g_1}(E,t)U_t\nonumber\\
 +(1+\ga_{g_1,g_0}(E)D)U_tB_{g_0}(E,t)\label{eq3}\\
 \fl \ga_{g_0,g_0}(E)D(1+\ga_{g_1,g_0}(E)D)B_{g_1}(E,t)U_t=
 -\ga_{g_0,g_0}(E)DD^+\ga_{g_1,g_1}(E)B_{g_0}(E,t)U_t\nonumber\\
 +\ga_{g_0,g_0}(E)DU_tB_{g_1}(E,t)\label{eq4}
\end{eqnarray}
Now after substitution of expression~(\ref{eq4}) into~(\ref{eq3})
one has
\begin{eqnarray}\fl
\Bigl(1+\ga_{g_0,g_1}(E)D^+-\ga_{g_1,g_0}(E)D+
(\ga_{g_0,g_0}\ga_{g_1,g_1}-\ga_{g_1,g_0}\ga_{g_0,g_1})(E)DD^+\Bigr)
B_{g_0}(E,t)U_t\nonumber\\
=\ga_{g_0,g_0}(E)DU_tB_{g_1}(E,t)+(1-\ga_{g_1,g_0}(E)D)U_tB_{g_0}(E,t)\label{eq5}
\end{eqnarray}
One can show by similar computations that
\begin{eqnarray}\fl
\Bigl(1+\ga_{g_0,g_1}(E)D^+-\ga_{g_1,g_0}(E)D+
(\ga_{g_0,g_0}\ga_{g_1,g_1}-\ga_{g_1,g_0}\ga_{g_0,g_1})(E)D^+D\Bigr)
B_{g_1}(E,t)U_t\nonumber\\
=-\ga_{g_1,g_1}(E)D^+U_tB_{g_0}(E,t)+(1+\ga_{g_0,g_1}(E)D^+)U_tB_{g_1}(E,t)\label{eq6}
\end{eqnarray}
Now since we suppose that the inverse operators $T_0(E)$ and $T_1(E)$ exist, we can solve the above
equations~(\ref{eq5}) and~(\ref{eq6}) with respect to $B_{g_0}(E,t)U_t$ and $B_{g_1}(E,t)U_t$. The solutions are
given by~(\ref{commbf}) and~(\ref{commbg}), and that proves the lemma.

Denote
\begin{eqnarray}
R_{0,0}(E):=\ga_{g_1,g_1}(E)DT_1(E)D^+\nonumber\\
R_{1,1}(E):=\ga_{g_0,g_0}(E)D^+T_0(E)D\nonumber\\
R_{0,1}(E):=-DT_1(E)(1+\ga_{g_0,g_1}(E)D^+)\nonumber\\
R_{1,0}(E):=D^+T_0(E)(1-\ga_{g_1,g_0}(E)D)\nonumber
\end{eqnarray}

\begin{theorem}
The normally ordered form of equation~(\ref{equ1}) is
\begin{equation}\label{normordeq1}
\frac{\rmd U_t}{\rmd t}=-\sum\limits_{n,m=0,1}\int\rmd
ER_{m,n}(E)B^+_{g_m}(E,t)U_tB_{g_n}(E,t)
\end{equation}
\end{theorem}
{\bf Proof.} Using~(\ref{H}) white noise Schr\"odinger
equation~(\ref{equ1}) can be rewritten in a more detailed form
\begin{equation}\label{equ21}
 \frac{\rmd U_t}{\rmd t}=\int\rmd E\bigl(D\om B^+_{g_0}(E,t)B_{g_1}(E,t)
 -D^+\om B^+_{g_1}(E,t)B_{g_0}(E,t)\bigr)U_t
\end{equation}
It follows from Lemma~\ref{normordu} that
\begin{eqnarray}
 D^+B_{g_0}(E,t)U_t=R_{1,1}(E)U_tB_{g_1}(E,t)+R_{1,0}(E)U_tB_{g_0}(E,t)\nonumber\\
 DB_{g_1}(E,t)U_t=-R_{0,0}(E)U_tB_{g_0}(E,t)-R_{0,1}(E)U_tB_{g_1}(E,t)\nonumber
\end{eqnarray}
The statement of the theorem is obtained after substitution of
these expressions in~(\ref{equ21}).

\begin{remark}
An immediate consequence of Theorem~2 is the following
factorization property of the limiting state $\varphi_L$:
\[
\varphi_L(B^+_{g_m}(E,t)U_tB_{g_n}(E,t))=
\varphi_L(B^+_{g_m}(E,t)B_{g_n}(E,t))\varphi_L(U_t)
\]
This property of the state $\varphi_L$ similar to the
factorization property of the state determined by a coherent
vector $\Psi, \|\Psi\|=1$:
\[
(\Psi,B^+_{g_m}(E,t)U_tB_{g_n}(E,t)\Psi)=
(\Psi,B^+_{g_m}(E,t)B_{g_n}(E,t)\Psi)(\Psi,U_t\Psi)
\]
which is usually used to define quantum stochastic differential
equations (the general notion of adaptedness and adapted domains
which are much larger than the coherent ones is given
in\,\cite{A}).
\end{remark}

Taking the partial expectation of both sides of equation~(\ref{normordeq1}) in the state $\varphi_L$, using the
factorization property and noticing that
\[
\varphi_L(B^+_{g_m}(E,t)B_{g_n}(E,t))=\langle g_n,P_ELg_m\rangle,
\]
one gets the equation
\begin{equation}\label{expU}
\frac{\rmd\varphi_L(U_t)}{\rmd t}=-\Gamma\varphi_L(U_t),
\end{equation}
where $\Gamma$ is being called drift and is equal to
\[
\Gamma = \sum\limits_{n,m=0,1}\int\rmd ER_{m,n}(E)\langle g_n,P_ELg_m\rangle
\]
The solution of~(\ref{expU}) is
\[
\varphi_L(U_t)=\rme^{-\Gamma t}
\]
In the case of orthogonal test functions, i.e. $\langle g_0,S_tg_1\rangle=0$ this expectation value for the
evolution operator was obtained in\,\cite{APV1}. Let us note that the expectation value is obtained in a
nonpertrubative way, without direct summation of the iterated series for the evolution operator, and is a result
of the procedure of causal normal ordering.

\section{ONE-PARTICLE $T$ OPERATOR AND $S$ MATRIX}\label{7}

In the low density limit the role of multiparticle collisions is negligible and the dynamics of the test particle
should be determined by the interaction of the test particle with one particle of the reservoir. In the present
section we give the expressions for the one-particle $T$-operator and $S$-matrix. In the next section we will
rewrite normally ordered white noise equation~(\ref{normordeq1}) in a form of the quantum stochastic
equation~(\ref{qsde1}) and show (Theorem 5) that the coefficients of this equation can be expressed in terms of
the one-particle $S$-matrix.

Because of number conservation, the closed subspace of $ {\cal H}_S\om\Gamma({\cal H})$ generated by vectors of
the form $ u\om A^+(f)\Phi $\enskip ($ u\in {\cal H}_{\rm S} $, $ f\in {\cal H}=L^2({\mathbb R}^d)$, $\Phi$ is the
vacuum vector), which is naturally isomorphic to $ {\cal H}_{\rm S}\om {\cal H} $, is globally invariant under the
time evolution operator $ \exp[\rmi (H_{\rm S}\om 1+1\om H_{\rm R}+V)t] $. The restriction of the time evolution
operator to this subspace corresponds to the evolution operator on $ {\cal H}_{\rm S}\om {\cal H}$ given by
\[
 \exp[\rmi(H_{\rm S}\om 1+1\om H_1+V_1)t]
\]
where
\begin{equation}\label{defV}
 V_1=\rmi(D\om|g_0\rangle\langle g_1|-h.c.)
\end{equation}

The one-particle M\o ller wave operators are defined as
\[
 \fl \Omega_{\pm}=s-\lim\limits_{t\to\mp\infty}\exp[\rmi(H_{\rm S}\om 1+1\om H_1+V_1)t]
 \exp[-\rmi(H_{\rm S}\om 1+1\om H_1)t]
\]
The one-particle $T$-operator is defined as
\begin{equation}\label{defT}
 T=V_1\Omega_+
\end{equation}
and the one-particle $S$-matrix as
\begin{equation}\label{defS}
 S=\Omega^*_-\Omega_+
\end{equation}

\begin{theorem}
For the interaction~(\ref{defV}) the one-particle $T$-operator and $S$-matrix have the form
\begin{equation}\label{Top}
T = -\rmi\sum\limits_{n,m\in\{0,1\}}\int\rmd ER_{m,n}(E)\om |g_m\rangle\langle P_Eg_n|
\end{equation}
\begin{equation}\label{Smat}
S = 1 -2\pi\sum\limits_{n,m\in\{0,1\}}\int\rmd ER_{m,n}(E)\om |P_Eg_m\rangle\langle P_Eg_n|
\end{equation}
\end{theorem}
{\bf Proof.} For the case $\langle g_0,S_tg_1\rangle =0$ equality~(\ref{Top}) was proved in~\cite{APV2}. The proof
of~(\ref{Top}) and~(\ref{Smat}) for the general case can be done in a similar way.

Expression~(\ref{Smat}) will be used in the next section for
derivation of equation~(\ref{qsde3}).

\section{QUANTUM STOCHASTIC EQUATION FOR THE LIMITING EVOLUTION OPERATOR}\label{s7}

Normally ordered white noise equation~(\ref{normordeq1})
equivalent, through identification
\[
B^+_m(E,t)U_tB_n(E,t)\rmd t=2\pi\rmd N_t(|P_Eg_m\rangle\langle P_Eg_n|)U_t
\]
to the quantum stochastic differential equation
\begin{equation}\label{qsde1}
\rmd U_t=-2\pi\sum\limits_{n,m\in\{0,1\}}\int\rmd ER_{m,n}(E)\rmd N_t(|P_Eg_m\rangle\langle P_Eg_n|)U_t
\end{equation}
where $N_t$ is the quantum Poisson process in $\Gamma(L^2(\mathbb
R_+)\om{\cal H})$ defined by $N_t(X):=N(\chi_{[0,t]}\om X)$, if
$X$ is an operator in ${\cal H}$. The stochastic differential
$\rmd N_t$ satisfies the usual Ito table
\begin{equation}\label{ito}
\rmd N_t(X)\rmd N_t(Y)=\rmd N_t(XY),
\end{equation}
where $X$, $Y$ are operators in ${\cal H}$, and the limit state
$\varphi_L$ characterized by the property
\[
\varphi_L(2\pi\rmd {\rm N}_t(|P_Ef\rangle\langle P_Eg|))=\langle g,P_ELf\rangle\rmd t
\]

The coefficients of quantum stochastic equation~(\ref{qsde1}) can be expressed in terms of one-particle $S$-matrix
describing scattering of the test particle on one particle of the reservoir. To show this we will use Hilbert
module notation. For any pair of Hilbert spaces $ {\cal X}_0,{\cal X}_1 $, if $N_t$ denotes the Poisson process on
the Fock space $ \Gamma(L^2(\mathbb R_+)\om{\cal X}_1) $, then for bounded operators $ X_0\in B({\cal X}_0) $, $
X_1\in B({\cal X}_1) $, the Hilbert module notation is\,\cite{FrMa}:
\[
 N_t(X_0\om X_1):=X_0\om N_t(X_1)
\]
With this notation equation~(\ref{qsde1}) can be written as
\begin{equation}\label{qsde2}
\rmd U_t=\rmd N_t\Bigl(-2\pi\sum\limits_{n,m\in\{0,1\}}\int\rmd ER_{m,n}(E)\om |P_Eg_m\rangle\langle
P_Eg_n|\Bigr)U_t
\end{equation}
An immediate conclusion from~(\ref{Smat}) and (\ref{qsde2}) is the
following theorem which is one of the main results of the paper.
\begin{theorem} The evolution operator in the low density limit satisfies
the quantum stochastic equation driven by the quantum Poisson process with intensity $S-1$:
\begin{equation}\label{qsde3}
\rmd U_t=\rmd {\rm N}_t(S-1)U_t
\end{equation}
\end{theorem}

Equation~(\ref{qsde3}) describes the dynamics of the compound system in the low density limit. Using this equation
and the Ito table for stochastic differentials one can obtain a quantum Langevin equation for the Heisenberg
evolution of any system observable. Then the corresponding master equation or, equivalently, quantum (linear)
Boltzmann equation for reduced density matrix of the system can be obtained simply by taking the partial
expectation of this Langevin equation in the causal state $\varphi_L$.

\begin{lemma}\label{unitarity} The solution of~(\ref{qsde3}) is unitary.
\end{lemma}
{\bf Proof.} Let us show that ${\rm d}(U^+_tU_t)=0$. The operator
$U^+_t$ satisfies the equation
\[
\rmd U^+_t=U^+_t\rmd N_t(S^+-1)
\]
One has
\begin{eqnarray}
{\rm d}(U^+_tU_t)={\rm d}U^+_tU_t+U^+_t{\rm d}U_t+{\rm d}U^+_t{\rm
d}U_t\nonumber\\
 =U^+_t{\rm d}N_t(S^+-1)U_t+U^+_t{\rm
d}N_t(S-1)U_t+ U^+_t{\rm d}N_t(S^+-1){\rm d}N_t(S-1)U_t\nonumber
\end{eqnarray}
Using the Ito table~(\ref{ito}) one gets
\[
{\rm d}N_t(S^+-1){\rm d}N_t(S-1)={\rm d}N_t((S^+-1)(S-1))
\]
This and unitarity of $S$ leads to
\[
{\rm d}(U^+_tU_t)=U^+_t{\rm d}N_t(S^+-1+S-1+(S^+-1)(S-1))U_t=0
\]
Now it follows from the initial condition $U_{t=0}=1$ that, for any $t\ge 0$, $U^+_tU_t=1$. The proof of
$U_tU^+_t=1$ can be done in a similar way.

\section{CONCLUSIONS}

In the present paper we consider the dynamics of a test particle ($N$-level atom) interacting with a dilute Bose
gas. It is proved that the dynamics of the total system converges in the low density limit to the solution of the
quantum stochastic equation driven by a quantum Poisson process with intensity $S-1$, where $S$ is the
one-particle scattering matrix. The limiting equation is derived in a nonpertrubative way, without use of iterated
series for the evolution operator. The derivation is based on the white noise approach and on the procedure of
causal normal ordering developed for the weak coupling limit by Accardi, Lu and Volovich\,\cite{ALV}. The novelty
of the present derivation is that it does not use the Fock-antiFock (or GNS) representation for the CCR algebra of
the Bose gas, determined by the state $\varphi_{L,\xi}$. This simplifies the derivation and allows us to express
the intensity of the Poisson process directly in terms of the one-particle $S$-matrix. The notion of causal states
is introduced and the convergence of the correlators of the free evolution of the initial number operators to
correlators of quantum white noise operators in causal states is proved. The causal states satisfy the
factorization property similar to that satisfied by states determined by coherent vectors. This property is
crucial for study of the reduced dynamics of the system.

\ack{The author is grateful to Professor L. Accardi and Professor Y.G. Lu for kind hospitality in the Centro Vito
Volterra and Bari University; to Professor L. Accardi, Professor Y.G. Lu, and Professor I.V. Volovich for many
useful and stimulating discussions. This work is partially supported by Grant INTAS YSF 01/1-200, a NATO-CNR
Fellowship, and Grant RFFI 02-01-01084.}

\section{APPENDIX: CAUSAL $\dl$-FUNCTION}
Let us recall the construction for distributions on the standard simplex (cf.~\cite{ALV}). Define
\[
 C_0:=\{\phi:\mathbb R_+\to\mathbb C\ |\ \phi=0 {\rm\ a.e.}\},
\]
\[
 C_1:=\{\phi:\mathbb R_+\to\mathbb C\ |\ \phi
 {\rm\ is\ bounded\ and\ left-continuous\ at\ any\ } t>0\},
\]
\[
 C:={\rm\ linear\ span\ of\ } \{C_0\cup C_1\}.
\]
For any $ a>0 $ define $\dl_+(\cdot-a) $ as the unique linear extension of the map:
\[
 \dl_+(\cdot-a):\phi\in C_1\to\phi(a)
\]
\[
 \dl_+(\cdot-a):\phi\in C_0\to 0.
\]
In~\cite{ALV} the following results are proved.

\begin{lemma}. In the sense of distributions one has the limit
\begin{equation}\label{dllimit}
 \lim\limits_{\la\to 0}{\frac{\rme^{\rmi(t'-t)E/\la^2}}{\la^2}} = 2\pi\dl(t'-t)\dl(E)
\end{equation}
\end{lemma}

\begin{lemma}. In the sense of distributions over the simplex $t\ge t'\ge 0$ one
has the limit
\begin{equation}\label{dlplimit}
 \lim\limits_{\la\to 0}{\frac{\rme^{\rmi(t'-t)E/\la^2}}{\la^2}} =\dl_+(t'-t)\frac{1}{\rmi (E-\rmi 0)}.
\end{equation}
\end{lemma}
The last equality means that for any $f\in C$, $g\in S(\mathbb R)$, one has the limit
\[
\lim\limits_{\la\to 0}\int\limits_0^t\rmd t'\int\limits_{\mathbb
R}\rmd E\, {\frac{\rme^{\rmi(t'-t)E/\la^2}}{\la^2}}f(t')g(E)
=f(t)\lim\limits_{\varepsilon\to 0+}\int\rmd E \frac{g(E)}{\rmi
(E-\rmi\varepsilon)}.
\]

\end{document}